\documentstyle[twocolumn,floats,aps]{revtex}
\begin{document}
\draft

\twocolumn[\hsize\textwidth%
\columnwidth\hsize\csname@twocolumnfalse\endcsname

\title{\bf Resonant Raman scattering by elementary electronic
excitations in semiconductor structures}

\author{S. Das Sarma and Daw-Wei Wang}
\address{Department of Physics, University of Maryland, College Park, 
MD 20742-4111}

\date{\today}
\maketitle
\pagenumbering{arabic}

\begin{abstract}
We explain quantitatively why resonant Raman scattering spectroscopy, an
extensively used experimental tool in studying elementary electronic
excitations in doped low dimensional semiconductor nanostructures, always
produces an observable peak at the so-called "single particle" excitation
although the standard theory predicts that there should be no such single
particle peak in the Raman spectra. We have thus resolved an experimental 
puzzle which dates back more than twenty-five years.
\end{abstract}


\vskip 1pc]
\narrowtext

\newcommand{\bfitk}{\mathbf{k}}
\newcommand{\bfitp}{\mathbf{p}}
\newcommand{\bfitq}{\mathbf{q}}

Resonant Raman scattering (RRS) involving inelastic scattering of light by
electrons has long been a powerful and versatile spectroscopic tool for
studying [1-10]
elementary excitations in doped low dimensional semiconductor structures such
as quantum wells, superlattices, and more recently, one dimensional quantum
wire systems. RRS has been extensively used in experimentally studying the
collective charge density excitation (CDE) dispersion in semiconductor
quantum wells,
quantum wires, and superlattices both for intrasubband and intersubband
transitions.
In the standard theory [11,12], which ignores the role of the valence
band and simplistically assumes the photon to be interacting entirely with
conduction band electrons, RRS intensity is proportional to the dynamical
structure factor [13] of the conduction band electron system and
therefore has peaks at the
collective mode frequencies at the appropriate wavevectors defined by the
experimental geometry. Restricting to the polarized 
RRS geometry [12], where the incident and scattered photons have the same
polarization vectors indicating the absence of spin flips in the electronic
excitations, the dynamical structure factor peaks should correspond to the
poles of the reducible density response function, which
are given by the collective CDEs of the system. In particular, the single
particle electron-hole excitations (SPE), which are at the poles of the
corresponding irreducible response function, 
carry no long wavelength spectral
weight in the density response function and should
\textit{not}, as a matter of principle, show up in the polarized RRS spectra.
The remarkable experimental fact, however, is that there is always
a relatively weak (but quite distinct) low energy SPE
peak in the observed RRS spectra (near resonance) in addition to the expected
strong CDE peak at higher energy. This observed SPE peak in the polarized
RRS spectra is a factor of $10^3$-$10^4$
times stronger than that given by the calculated dynamical structure factor
in the standard theory. This
puzzling feature of an ubiquitous anomalous SPE peak (in addition to
the expected CDE peak) in the observed RRS spectra occurs in one dimensional
GaAs-AlGaAs quantum wires, two dimensional GaAs quantum wells, and even in
the doped three dimensional bulk GaAs systems
[1]. It exists in the low dimensional structures both for intrasubband and
intersubband (i.e. transitions along and perpendicular to the confining
quantization direction) excitations, and in zero and finite magnetic fields.
No theoretical understanding of this phenomenon exists in spite of the great
ubiquitousness of the effect. Ad hoc proposals [2,14] have been made
in the literature that perhaps a serious breakdown of momentum or wavevector
conservation (arising, for example, from scattering by random impurities) is
responsible for somehow transferring spectral weight from large to small
wavevectors. Apart from being completely ad hoc, this proposal also suffers
from any
lack of empirical evidence in its support --- in particular, the observed
anomalous SPE peak in the RRS spectra does not correlate with the
strength of the impurity scattering in the system. The ubiquitousness of the
phenomenon suggests that it must arise from some generic principle underlying
RRS itself, and cannot be explained by non-generic and
manifestly system-specific proposals which have been made occasionally
in the literature. 
We emphasize that this very basic lack of understanding of why an SPE peak 
shows up in the experimental RRS spectra of doped semiconductor structures 
is an important problem because the RRS is one of the most powerful 
techniques to study an interacting electron system in one, two and three 
dimensions, including even quantum Hall systems.

In this article we provide a quantitative and
\textit{compellingly generic} theoretical explanation for this puzzle.
We emphasize that
the generic nature of the phenomenon strongly suggests that its quantitative
explanation must lie in the fundamental principles of RRS and must not
depend on the experimental details [1-10], such as the system dimensionality,
intra- or inter-subband excitations being probed, the existence (or not) of
an external magnetic field, etc. Our theory depends only on the resonant nature
[12,15-17] of the experiment (i.e. the laser frequency of the external light
used in the RRS is approximately equal to the fundamental band gap of the
semiconductor, which causes a resonant enhancement of the RRS intensity
allowing the observation of the elementary electronic excitations in the
conduction band which usually do not couple to light).

In Fig. 1(a) we depict the schematic diagram [15-17] for the
two step process (steps 1 and 2 in the figure) involved in the 
polarized RRS spectroscopy at the
$E_0+\Delta_0$ direct gap of GaAs [17] where an electron in the valence
band is excited by the incident photon into an excited (i.e. above the Fermi
level) conduction band state, leaving a valence band hole behind (step 1);
then an electron from inside the conduction band Fermi surface combines with
the hole in the valence band (step 2).
Electron spin is conserved throughout the scattering
process. 
\begin{figure}

 \vbox to 5.cm {\vss\hbox to 6cm
 {\hss\
   {\includegraphics{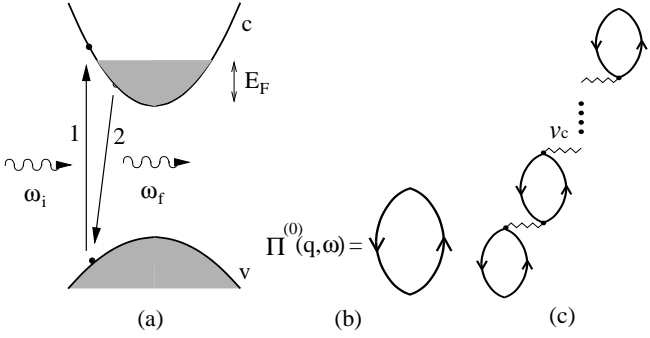}
   }
  \hss}
 }
\caption{(a) Schematic representation of the resonant Raman scattering in the
doped, direct gap two band model of GaAs nanostructure. $\omega_i$ and
$\omega_f$ are the initial and final frequencies of the external photon.
Steps 1 and 2 are described in the text. The RRS is a two step process
involves steps 1 and 2, leaving an excited electron-hole pair in the conduction 
band. (b) Diagrammatic representation of
the conduction band irreducible polarizability,
$\Pi^{(0)}(\bfitq,\mathit{\omega})$, in RPA calculation.
(c) Diagrammatic representation of the conduction band reducible polarizability,
$\Pi(\bfitq,\mathit{\omega})$, in the standard theory.
$v_c(\bfitq)$ is the Coulomb interaction.
}
\end{figure}
In the standard random phase approximation (RPA), which neglects
all interband resonance
effects and considers only the conduction band,
the RRS intensity will then be proportional [11] to the imaginary
part of the reducible response function, which is given by
$-$Im$\Pi (\bfitq,\omega)$, where $\Pi(\bfitq,\mathit{\omega})=
\mathrm{\Pi^{(0)}}
(\bfitq,\mathit{\omega})\,
\epsilon(\bfitq,\mathit{\omega})^{-1}$ is the reducible polarizability with
$\Pi^{(0)}(\bfitq,\omega)$ as the non-interacting conduction band
electron polarizability and $\epsilon(\bfitq,\omega) = 1-\mathit{v_c}(\bfitq)
\,\mathrm{\Pi^{(0)}}(\bfitq,\omega)$ is the appropriate
dynamical dielectric function (with $v_c(\bfitq)$ as the Coulomb interaction)
--- the
geometric series of "bubble" diagrams in Fig. 1(c) is the characteristic
feature of a
CDE or a plasmon excitation. We show in Fig. 2 some typical calculated one
dimensional (1D) RRS spectra based on this simple formula 
which has been universally employed [11] in the
literature. The calculated spectrum gives reasonable quantitative agreement
with the experimentally observed [6] polarized RRS spectra for the CDE
peak in 1D GaAs quantum wires (the same is true in higher dimensions --- see,
for example, refs [11]) except for one important feature --- the theoretical
spectrum has only one peak corresponding to the CDE whereas experimentally one
always sees an SPE peak at resonance.

\begin{figure}

 \vbox to 5.5cm {\vss\hbox to 5.cm
 {\hss\
   {\includegraphics{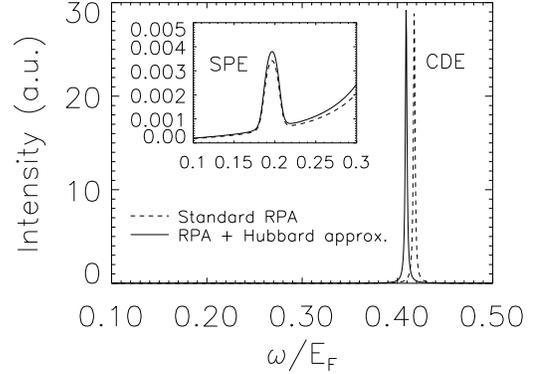}
   }
  \hss}
 }
\caption{
The dynamical structure factor in the standard theory
for 1D quantum wire system calculated by RPA
in long wavelength limit ($q=0.1k_F$). Vertex correction by
Hubbard approximation is also shown for comparison.
The SPE peak (inset) is much smaller ($\sim 10^{-4}$) than the CDE peak in
the standard theory.
Finite impurity scattering effect (included in the theory) leads to the
broadening of the peaks.
}
\end{figure}
We now consider the full resonance situation including the
valence band which obviously [12,15-18] plays a crucial role in the RRS
experiment because the external photon energy must approximately equal the
$E_0+\Delta_0$ direct gap for the experiment to succeed. We can write
(assuming the
usual $\bfitp\cdot \mathbf{A}$ coupling of light to matter)
the RRS scattering cross
section [17] for the resonant scattering process for the conduction
band electrons as (using $i$, $f$
to denote the initial pre-scattering and the final post scattering states):
\begin{equation}
\frac{d^2\sigma}{d\Omega d\omega}\propto \frac{\omega_f}{\omega_i}
\langle\Sigma_{F} |M_{FI}|^2\delta (E_F-E_I-\omega)\rangle_I,
\end{equation}
where $\omega_f$, $\omega_i$ are the final and initial photon energies
, and $\omega$ is their difference ($\hbar=1$ and
a sum over all final electronic states, $F$, and an average over all initial
electronic states, $I$, are implied in Eq. (1)).
$M_{FI}$ is the transition matrix element defined by
(using the subscripts $c$ and $v$ to denote the conduction and valence band
electronic Bloch states respectively in our two band model) [19]
\begin{figure}

 \vbox to 5.5cm {\vss\hbox to 5.cm
 {\hss\
   {\includegraphics{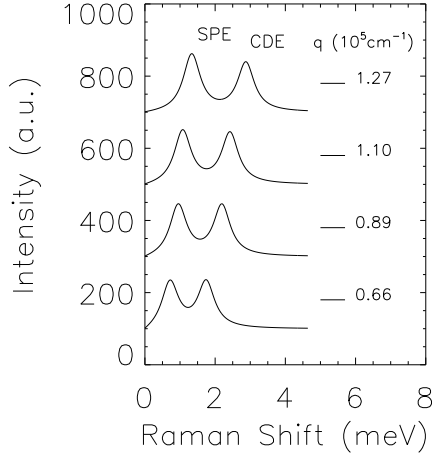}
   }
  \hss}
 }
\caption{
The calculated resonant Raman scattering intensity
in the full theory
including the interband transition for a realistic 1D quantum wire
system. The SPE spectral weight is enhanced by the resonance
and is now comparable to that of CDE. The parameters are chosen to correspond
to the experimental system of ref. [6], where the resonance condition
is satisfied. Our theoretical results agree very well with the experimental
data shown in the Fig. 1 of ref. [6].
}
\end{figure}
\begin{eqnarray}
M_{FI}& = & \langle F|\Sigma_{\bfitk}\mathit{\gamma(\bfitk)}
c^{\dagger}_{\bfitk}c_{\bfitk-\bfitq}|I \rangle \nonumber \\
\mathit{\gamma(\bfitk)}& = &\mathit{\hat{n}_i\cdot\hat{n}_f} \\
 & + & \;\;
\frac{\mathrm{1}}{m_c}
\left( \frac{ _c\langle\bfitk | \hat{\mathit{n_f}}\cdot\bfitp |
\bfitk+\bfitk_{\mathit{f}}
\rangle _{\mathit{v}} \;
_{\mathit{v}}\langle\bfitk+\bfitk_{\mathit{f}} | \hat{\mathit{n}}
_{\mathit{i}}\cdot\bfitp | \bfitk-\bfitq\rangle_{\mathit{c}}}
{E_g+E_c(\bfitk)-\mathit{E_v}(\bfitk+\bfitk_{\mathit{f}})
+\mathit{\omega_i}} \right. \nonumber \\
 & + & \;\;
\left. \frac{ _c\langle\bfitk | \hat{\mathit{n_i}}\cdot\bfitp | \bfitk
-\bfitk_{\mathit{i}}\rangle _{\mathit{v}} \;
_{\mathit{v}}\langle\bfitk-\bfitk_{\mathit{i}} | \hat{\mathit{n}}
_{\mathit{f}}\cdot\bfitp | \bfitk-\bfitq\rangle_{\mathit{c}}}
{E_g+E_c(\bfitk)-\mathit{E_v}(\bfitk-\bfitk_{\mathit{i}})
-\mathit{\omega_i}} \right)
\end{eqnarray}
where $c_{\bfitk}$'s are Fermion operators of wavevector $\bfitk$ and
the spin index is neglected;
$\hat{n}_i/\hat{n}_f$, $\bfitk_{\mathit{i}}/\bfitk_{\mathit{f}}$,
$\omega_i/\omega_f$, are respectively the
initial/final photon polarization vectors, wavevectors and frequencies;
$|\bfitk\rangle_{\mathit{c,v}}$ and
$E_{c,v}(\bfitk)$ refer to conduction/valence band Bloch states
and energies for a wavevector $\bfitk$ in the corresponding band;
$\omega\equiv\omega_i-\omega_f$, $\bfitq\equiv\bfitk_{\mathit{i}}-
\bfitk_{\mathit{f}}$ are
the energy and wavevector difference of the photons in the experiment.
Assuming the resonance condition, i.e. $\omega_i\approx E_g+E_c(k_F)-E_v(k_F)$,
where $E_g$ is
the $E_0 +\Delta_0$ band gap and $k_F$ is the Fermi wavevector in the
conduction
band; the usual back
scattering experimental geometry, i.e. $\bfitq=\mathrm{2}\bfitk_{\mathit{i}}$;
and the
well-satisfied condition $|\bfitk_{\mathit{i}}|\sim|\bfitk_{\mathit{f}}|\ll
\mathit{k_F}$ due to the
long wavelength ($\sim$ 5000 \AA) of visible light compared with the Fermi
wavelength ($\sim$ 100 \AA$\,$or less), Eqs. (1) and (2) can be written as
\begin{equation}
\frac{d^2\sigma}{d\Omega d\omega}\sim -\mathrm{Im}\left[\Pi^{(2)}(\bfitq,\omega)
+
\frac{\left[\mathrm{\Pi^{(1)}}(\bfitq,\omega)\right]^{\mathrm{2}}{\mathit{v_c}}
(\bfitq)}
{\epsilon(\bfitq,\omega)}
\right],
\end{equation}
where
\begin{equation}
\Pi^{(n)}(\bfitq,\mathit{\omega})=\frac{-\mathrm{2}}{(\mathrm{2}
\mathit{\pi})^D}\int d^Dp\;
\frac{[A(\bfitp,\bfitq)]^{\mathit{n}}(\mathit{n_c}({\bfitp})-\mathit{n_c}(
{\bfitp-\bfitq}))}
{\omega +i\delta -E_c({\bfitp})+E_c({\bfitp-\bfitq})}, \nonumber \\
\end{equation}
where $D$ is the dimensionality, and $n_c(\bfitp)$ refers to the conduction
band Fermi occupancy for wavevector $\bfitp$ with
\begin{equation}
A(\bfitp,\bfitq)=[\mathit{E_{\omega}}+(\mathrm{1}+\xi)(\tilde{\bfitp}^2-
\mathrm{1})+\xi
(-\tilde{\bfitp}\cdot\tilde{\bfitq}+\tilde{\bfitq}^{\mathrm{2}}/4)]
^{-\mathrm{1}}, \nonumber \\
\end{equation}
where $E_{\omega}\equiv {E_F}^{-1}(E_g+(1+\xi)
{k_{F}}^2 /2m_c-\omega_i)$;
$\xi\equiv m_c/m_v$; $\tilde{\bfitp}\equiv \bfitp/\mathit{k_F}$;
$\tilde{\bfitq}\equiv \bfitq/\mathit{k_F}$, and $E_F=E_c(k_F)$ is the Fermi
energy of the conduction band electrons.
We have assumed parabolic band dispersions near the band extrema with $m_c$ and
$m_v$ as the conduction and valence band effective masses. Using GaAs band
parameters we can now calculate the polarized RRS spectra from Eq. (3).
Note that the resonance effects are nonperturbative and depend crucially
on the exact value of the incident photon energy.
Our theory can be considered to be a \textit{resonant} RPA theory which
explicitly takes into account the interband resonant process involved in the
RRS experiments.
\begin{figure}

 \vbox to 5.5cm {\vss\hbox to 5.cm
 {\hss\
   {\includegraphics{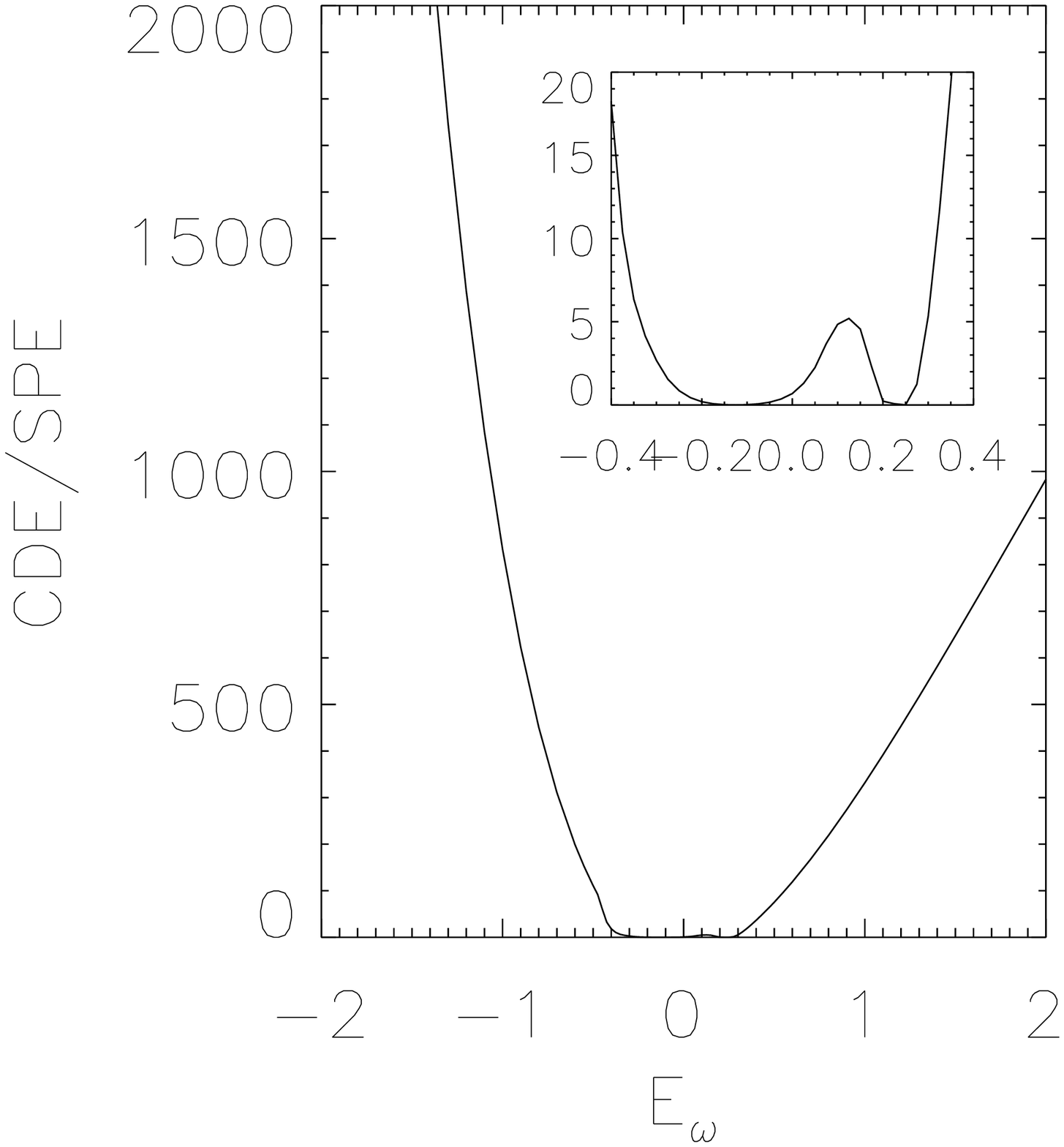}
   }
  \hss}
 }
\caption{
The resonant Raman scattering spectral weight ratio of CDE to SPE as a function
of the resonance parameter, $E_{\omega}$. For off-resonance,
$|E_\omega|\geq 0.5$, CDE always dominates SPE, but within the
resonance region, $|E_\omega|<0.5$, the SPE could be stronger than CDE as
shown in the inset. All parameters are the same as those in Fig. 3.
}
\end{figure}

In Fig. 3-5 we give our representative results for the
calculated polarized RRS spectra for intrasubband elementary excitations for
GaAs 1D and 2D structures. In Fig. 3 the specific wavevectors and other
system details have been chosen for the experimental GaAs quantum wires of
ref. [6]. Our calculated spectra at resonance are in quantitative
agreement with the corresponding experimental RRS spectra shown in Fig. 1 of
ref.[6]. The calculated SPE spectral weight at resonance in our Fig. 3 is
comparable to that of the CDE (to be contrasted with the simple non-resonant
calculation shown in Fig. 2). We emphasize that this spectacular
enhancement of SPE spectral weight in the resonant scattering
process disappears as one moves away from resonance. This can
be seen in Fig. 4,
\begin{figure}
 \vbox to 5.cm {\vss\hbox to 5.cm
 {\hss\
   {\includegraphics{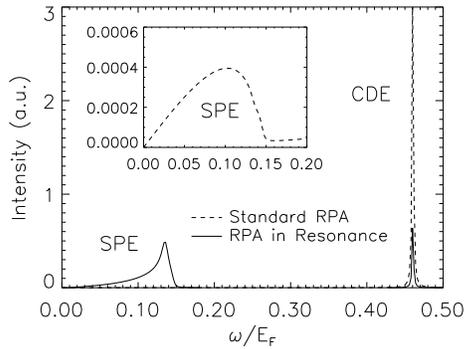}
   }
  \hss}
 }
\caption{
The calculated 2D resonant Raman scattering intensity in the full theory
in the long wavelength limit ($q=0.1k_F$). Simple
RPA result in the standard theory is also shown for comparison.
The resonance effect ($E_{\omega}=0.23$) strongly enhances SPE and suppresses
CDE in the 2D system similar to the 1D system. The inset shows the very
weak SPE peak in the standard theory, which should be manifestly experimentally
unobservable.
}
\end{figure}
where we plot the spectral weight ratio of
CDE/SPE weight as a function of the resonance condition itself. It is obvious
that the SPE spectral weight, while having a rather nontrivial structure
around the resonance condition, is essentially zero far away from resonance
where the CDE dominates. Experimentally it
is well-known that the SPE spectral weight dies off as the
incident photon energy goes off-resonance. Our detailed theoretical results
provide specific nontrivial predictions about how the SPE spectral weight
should vary as a function of the incident photon energy. Finally in Fig. 5
we show some representative calculations for 2D polarized RRS spectra (as
appropriate for elementary conduction band excitations in GaAs quantum wells)
in 2D systems.
While the quantitative details for the 2D systems differ from the
corresponding 1D results the basic theoretical phenomenon is the same: the
SPE spectral weight is anomalously enhanced at resonance compared with the
simple RPA, whereas off-resonance the SPE spectral weight decreases, eventually
becoming negligiblly small.

As a concluding note it may be important to emphasize that the nomenclature 
"SPE" peak, which we have used through this paper following the standard 
experimental literature [1-10], characterizing the low energy RRS peak is 
inappropriate since the SPE strictly corresponds to a peak in Im$\Pi^{(0)}$. 
We also note that interaction effects have been neglected in our 
irreducible response function in the spirit of RPA, which is entirely 
justifiable in two and three dimensions, but is open to question in 1D. We are 
currently [20] investigating 1D interaction effects on the RRS spectra by going 
beyond the resonant RPA scheme of Eqs. (1)-(6) --- we find that perturbative 
or mean-field (e.g. Hubbard approximation) inclusion of interaction effects 
does not qualitatively affect the RPA results. The striking phenomenological 
similarity in the experimentally observed RRS spectra in 1-, 2-, and 
3-dimensional systems is a strong indication that generic resonance physics 
as studied in this paper (within a \textit{resonant} RPA scheme) is playing 
a fundamental role in producing the low energy "SPE" feature in the polarized 
RRS spectra. 

We thank A. J. Millis for critical discussions, and the US-ONR and the US-ARO
for support.


\begin{thebibliography}{23}
\bibitem{1} A. Pinczuk \textit{et al}, 
             Phys. Rev. Lett. \textbf{27}, 317 (1971).
\bibitem{2} A. Pinczuk \textit{et al},
             Phys. Rev. Lett. \textbf{61}, 2701 (1988).
\bibitem{3} A. Pinczuk \textit{et al},
             Phys. Rev. Lett. \textbf{63}, 1633 (1989).
\bibitem{4} D. Gammon \textit{et al}, 
             Phys. Rev. B \textbf{41}, 12311 (1990);
            M. Berz \textit{et al}, 
             Phys. Rev. B \textbf{42}, 11957 (1990).
\bibitem{5} A. Pinczuk \textit{et al},
             Phys. Rev. Lett. \textbf{70}, 3983 (1993).
\bibitem{6} A. R. Go$\tilde{\mathrm{n}}$i \textit{et al}, 
             Phys. Rev. Lett. \textbf{67}, 3298 (1991).
\bibitem{7} D. Gammon \textit{et al},
             Phys. Rev. Lett. \textbf{68}, 1884 (1992);
            R. Decca \textit{et al}, 
             Phys. Rev. Lett. \textbf{72}, 1506(1996).
\bibitem{8} A. Schmeller \textit{et al}, 
             Phys. Rev. B \textbf{49}, 14778 (1994).
\bibitem{9} C. Sch$\ddot{\mathrm{u}}$ller \textit{et al},
             Phys. Rev. B \textbf{54}, R17304 (1996).
\bibitem{10} R. Strenz \textit{et al}, 
              Phys. Rev. Lett. \textbf{73}, 3022 (1994).
\bibitem{11} J. K. Jain and P. B. Allen,
              Phys. Rev. Lett. \textbf{54}, 947 (1985);
             J. K. Jain and P. B. Allen,
              Phys. Rev. Lett. \textbf{54}, 2437 (1985);
             S. Das Sarma and E. H. Hwang,
              Phys. Rev. Lett. \textbf{81}, 4216 (1998).
\bibitem{12} A. Pinczuk \textit{et al},
              Philosophical Magazine B \textbf{70}, 429 (1994) and
              references therein.
\bibitem{13} D. Pines, \textit{The Theory of Quantum Liquids}
              (Benjamin, New York, 1966).
\bibitem{14} I. K. Marmorkos and S. Das Sarma
              Phys. Rev. B \textbf{45}, 13396 (1992)
\bibitem{15} M. V. Klein, 
              p. 147 in \textit{Light Scattering in Solids},
              edited by Cardona, M. (Springer-Verlag, New York, 1975).
\bibitem{16} E. Burstein \textit{et al}, 
              Surf. Sci. \textbf{98}, 451 (1980).
\bibitem{17} A. Pinczuk and G. Abstreiter,
              p. 153 in \textit{Light Scattering in Solids V}, edited by
              Cardona, M. and G$\ddot{\mathrm{u}}$ntherodt, G.
              (Springer-Verlag, New York, 1989).
\bibitem{18} P. A. Wolff,
              Phys. Rev. \textbf{171}, 436 (1968);
             F. A. Blum, 
              Phys. Rev. B \textbf{1}, 1125 (1970).
\bibitem{19} P. A. Wolff, 
             Phys. Rev. Lett. \textbf{16}, 225 (1966).
\bibitem{20} D. W. Wang, A. J. Millis, and S. Das Sarma, unpublished.

\end{thebibliography}
\end{document}